\documentclass{ws-p8-50x6-00}

\begin{document}
\newcommand {\epem} {e$^+$e$^-$}
\newcommand {\qq} {q\overline{q}}
\newcommand {\gincl} {$g_{incl.}$}
\newcommand {\ecm} { E_{c.m.} }
\newcommand {\evis} { E_{vis.} }
\newcommand {\nqqg} { {\mathrm{N}}_{q\overline{q}g} }
\newcommand {\nqqgch} { {\mathrm{N}}_{q\overline{q}g}^{ch.} }
\newcommand {\nqq} { {\mathrm{N}}_{q\overline{q}} }
\newcommand {\ngg} { {\mathrm{N}}_{gg} }
\newcommand {\ngluon} { {\mathrm{N}}_{g} }
\newcommand {\nq} { {\mathrm{N}}_{q} }
\newcommand {\nqqch} { {\mathrm{N}}_{q\overline{q}}^{ch.} }
\newcommand {\nggch} { {\mathrm{N}}_{gg}^{ch.} }
\newcommand {\nchee} { {\mathrm N}^{\,ch.}_{ 
           {\mathrm e}^+{\mathrm e}^- } }
\newcommand {\lqq} { {L}_{q\overline{q}} }
\newcommand {\ktle} { k_{\perp,Le} }
\newcommand {\ktlu} { k_{\perp,Lu} }
\newcommand {\ptle} { p_{\perp,Le} }
\newcommand {\ptlu} { p_{\perp,Lu} }
\newcommand {\kperp} { k_{\perp} }
\newcommand {\ycut} { y_{cut} }
\newcommand {\dsign} { d_{sign} }
\newcommand {\caa} { C_{\mathrm A} }
\newcommand {\cff} { C_{\mathrm F} }
\newcommand {\nff} { n_f }
\newcommand {\sqq} { s_{q\overline{q}} }

\title{Unbiased gluon jet multiplicity from
{\epem} three-jet events
}

\author{J. William GARY}

\address{Department of Physics, University of California,
Riverside, CA 92521, USA\\ 
E-mail: bill.gary@ur.edu}

\maketitle

\abstracts{
The charged particle multiplicities of two- and
three-jet events from the reaction
{\epem}$\,\rightarrow\,$Z$^0$$\,\rightarrow\,$$hadrons$
are measured for Z$^0$ decays to light quark (uds) flavors,
using the data sample of the OPAL Collaboration at LEP.
Using recent theoretical expressions to account for
biases from event selection,
results corresponding to unbiased gluon jets are extracted.
The unbiased gluon jet data
are compared to corresponding results for quark jets.
We determine
the ratio $r$$\,\equiv\,$$\ngluon/\nq$
of multiplicities between
gluon and quark jets as a function of energy scale.
We also determine the ratio of slopes,
$r^{(1)}$$\,\equiv\,$$({\mathrm{d}}\ngluon /{\mathrm{d}}y)
/({\mathrm{d}}\nq / {\mathrm{d}}y)$,
and of curvatures,
$r^{(2)}$$\,\equiv\,$$({\mathrm{d}}^2\ngluon /{\mathrm{d}}y^2)
/({\mathrm{d}}^2\nq / {\mathrm{d}}y^2)$,
where $y$ specifies the energy scale.
At 30~GeV,
we find $r$$\,=\,$$1.422\pm0.051$,
$r^{(1)}$$\,=\,$$1.761\pm0.071$ and
$r^{(2)}$$\,=\,$$1.98\pm0.13$,
where the uncertainties are the statistical and systematic
terms added in quadrature.
These results are in general agreement with theoretical
predictions.
}

\section{Introduction}

The mean charged particle multiplicity of a gluon jet
has often been measured in the annihilation of an 
electron and positron to hadrons,
{\epem}$\,\rightarrow\,$$hadrons$.
The usual method
is to select three-jet quark-antiquark-gluon $q\overline{q}g$ 
final states for which the events and individual jets 
are defined using a jet algorithm.
The particle multiplicity of a jet 
determined with this technique
is found to depend on which algorithm is employed.
Therefore, these jets and the associated $q\overline{q}g$ 
events are called ``biased.''
In contrast,
theoretical calculations usually define gluon
jet multiplicity inclusively,
by the particles in hemispheres of 
gluon-gluon ($gg$) systems in an overall color singlet.
Quark jets are defined analogously as hemispheres
of quark-antiquark ($q\overline{q}$) systems.
The hemisphere definition of jets yields results which
are independent of a jet finder.
Therefore, these jets are called ``unbiased.''
Unbiased gluon jet multiplicity has so far been measured 
only in $\Upsilon$~\cite{bib-cleo92,bib-cleo97} and
Z$^0$~\cite{bib-opalgincl96,bib-opalgincl97,bib-opalgincl98} decays,
corresponding to jet energies of about
5 and 40~GeV, respectively.

It is of interest to measure the multiplicity
of unbiased gluon jets at other scales.
Such measurements would allow a test of 
recent predictions~\cite{bib-lupia,bib-capella}
from Quantum Chromodynamics (QCD)
for the scale dependence of the multiplicity 
ratio $r$$\,\equiv\,$$\ngluon/\nq$ 
between gluon and quark jets and for
the related ratios of slopes 
$r^{(1)}$$\,\equiv\,$$({\mathrm{d}}\ngluon /{\mathrm{d}}y)$ and
of curvatures 
$r^{(2)}$$\,\equiv\,$$({\mathrm{d}}^2\ngluon /{\mathrm{d}}y^2)$,
where $\ngluon$ and $\nq$ are the mean particle multiplicities
of gluon and quark jets,
$y$$\,=\,$$\ln\,(Q/\Lambda)$,
$Q$ is the jet energy and $\Lambda$ is the QCD scale parameter.
Recently,
a method to extract the multiplicity of unbiased
gluon jets from {\it biased} 
{\epem}$\,\rightarrow\,$$q\overline{q}g$ 
events was proposed~\cite{bib-eden,bib-edenkhoze},
extending earlier formalism~\cite{bib-dok88}.
By combining measurements of $\ngluon$ found from this method
with the unbiased measurements for $\nq$,
the ratios $r$, $r^{(1)}$ and $r^{(2)}$ can be determined
at a variety of scales and used to test the corresponding
QCD results in a quantitative manner.

Note that the DELPHI Collaboration has also recently
presented preliminary results testing the theoretical
formalism of~\cite{bib-eden,bib-edenkhoze}
(see~\cite{bib-delphiprelim}).

\section{Analysis method}

Analytic expressions for the mean particle multiplicity
of {\epem} three-jet events,
valid to the next-to-leading order of perturbation
theory (NLO, also called MLLA),
were recently presented in~\cite{bib-edenkhoze}:
\begin{eqnarray}
  \nqqg & = & \nqq\,(L,\ktlu) + \frac{1}{2}\, \ngg\,(\ktlu)
 \label{eq-eden14b} \;\;\;\; , \\
  \nqqg & = & \nqq\,(\lqq,\ktlu) + \frac{1}{2}\, \ngg\,(\ktle) 
 \label{eq-eden14a}
   \;\;\;\; .
\end{eqnarray}
The reason for the two different expressions
is that there is an ambiguity in the definition of the gluon jet
transverse momentum when the gluon radiation is hard.
$\ktlu$ and $\ktle$ are the transverse momenta of the
gluon with respect to the quark-antiquark system using
the definition of either the Lund ($\ktlu$)~\cite{bib-eden}
or Leningrad ($\ktle$)~\cite{bib-dok88} groups.
$\nqqg$ is the particle multiplicity of a three-jet event
selected using a jet algorithm.
$\nqq$ and $\ngg$ are the multiplicities of two-jet
$q\overline{q}$ and $gg$ systems,
given by twice $\nq$ and $\ngluon$, respectively.
The scales $L$, $\lqq$, $\ktlu$ and $\ktle$ 
are defined in~\cite{bib-eden,bib-edenkhoze}.

The multiplicity of the $gg$ system in this formalism,
$\ngg$,
depends only on a single scale:
$\ktlu$ in eq.~(\ref{eq-eden14b}) or
$\ktle$ in eq.~(\ref{eq-eden14a}).
This dependence on a single scale is a statement
that $\ngg$ is {\it unbiased},
i.e. $\ngg\,(\kperp)$ in eq.~(\ref{eq-eden14b})
or~(\ref{eq-eden14a}) is equivalent to the inclusive
multiplicity of a $gg$ event from a color singlet source 
produced at the same scale~$\kperp$,
to NLO accuracy.

The theoretical formalism is based on massless quarks.
Therefore,
we select light quark (u, d and~s) events for our study.
Three-jet events are defined using a jet algorithm.
For our standard analysis
we employ the Durham jet finder.
As a systematic check,
we use the Cambridge and Luclus jet finders.
The resolution scale of the Durham jet finder, $\ycut$,
is adjusted separately for each tagged uds event so that
exactly three jets are reconstructed.
For the Cambridge jet finder,
the resolution scale is again~$\ycut$.
For Luclus the corresponding parameter is~$d_{join}$.
The jets are ordered from 1 to 3
such that jet~1 has the highest energy.
The angle opposite jet~1 is called $\theta_1$, etc.
Events are retained if the angles between the highest
energy jet and the other two are the same to within~3$^\circ$,
the so-called ``Y events''~\cite{bib-opalse91,bib-opalqg91}.
For Y events,
the three-jet event multiplicity and scales
$\lqq$, $\ktlu$ and $\ktle$,
depend only on $\ecm$ and one inter-jet angle,
conveniently chosen to be~$\theta_1$.
For $35^{\circ}$$\,\leq\,$$\theta_1$$\,\leq\,$$120^{\circ}$,
the range of $\theta_1$ we employ for our 
gluon jet analysis,
$22\,365$ events are selected:
this is our final event sample.
For simplicity,
we identify the gluon jet by assuming it is the 
lowest energy jet in an event, i.e. jet~3.

To find the $\nqq\,(L,\ktlu)$ terms in eq.~(\ref{eq-eden14b}),
we employ two methods.
First, 
for the standard analysis,
we perform a direct measurement.
Specifically we determine the particle multiplicity of two-jet 
uds events from Z$^0$ decays 
as a function of the jet resolution scale~$\ktlu$.
Second, 
as a systematic check,
we evaluate the following analytic expression
from~\cite{bib-eden}:
\begin{equation}
  \nqq\,(J,\ktlu) \; = \; \nqq\,(J^{\prime})
     + (J - J^{\prime})\, \frac{{\mathrm{d}}\,\nqq\,(J^{\prime}) }
        {{\mathrm{d}}\,J^{\prime} } \;\;\;\; ,
  \label{eq-qbiased}
\end{equation}
where $J^{\prime}$$\,=\,$$\ktlu+c_q$
with $c_q$$\,=\,$3/2.
The unbiased term $\nqq\,(K)$ is equivalent to the
mean multiplicity of inclusive
{\epem}$\,\rightarrow\,$$hadrons$ events 
as a function of $K$$\,=\,$$\ecm$.
To find the $\nqq\,(\lqq,\ktlu)$ terms
in eq.~(\ref{eq-eden14a}),
we utilize only the second of these methods,
i.e. the one based on eq.~(\ref{eq-qbiased}),
beacuse a direct measurement is not straightforward in this case.

\section{Results}

\begin{figure}[t]
\begin{center}
  \begin{tabular}{ccc}
    \epsfxsize=8.5pc \epsffile{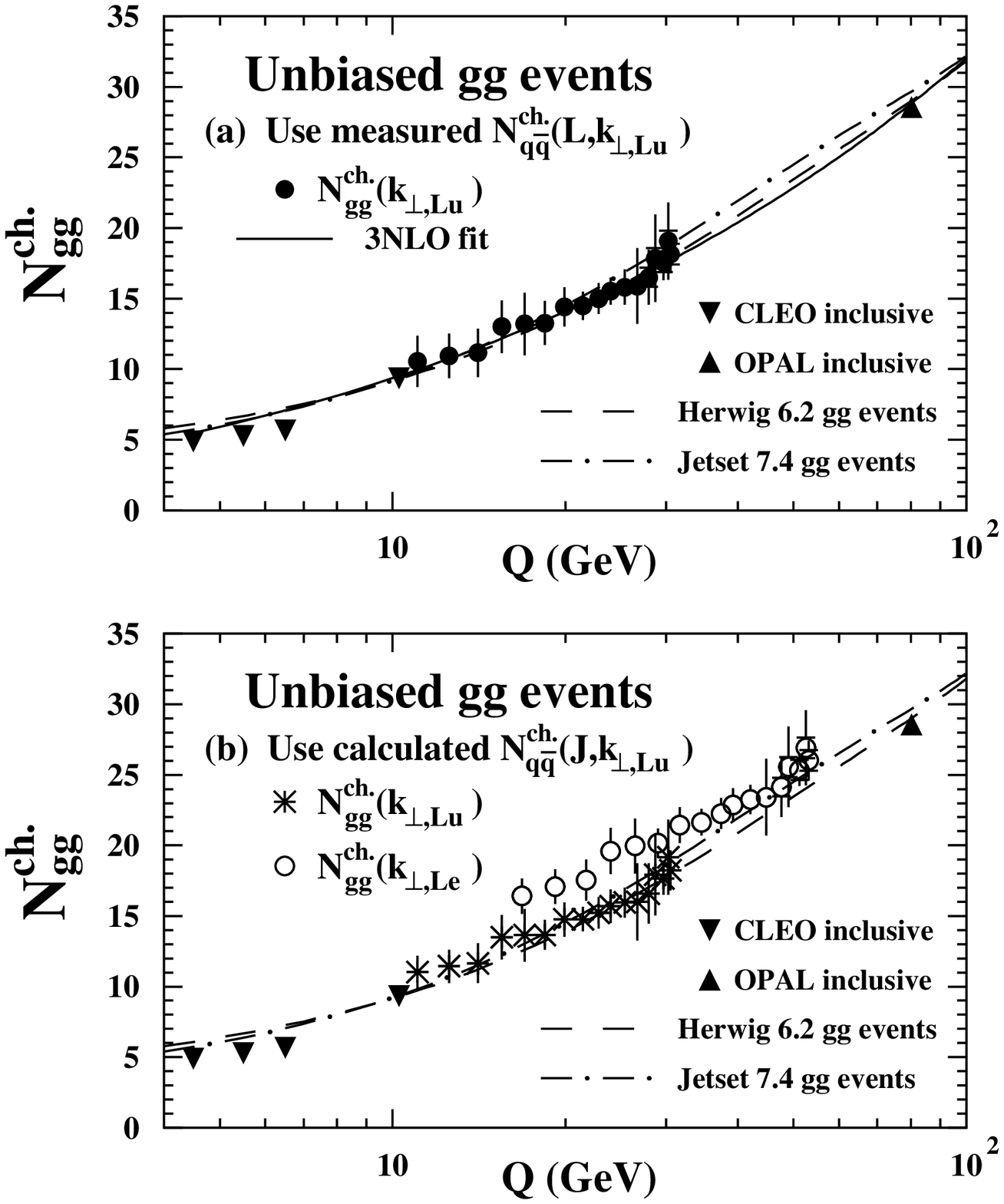} &
    \epsfxsize=8.5pc \epsffile{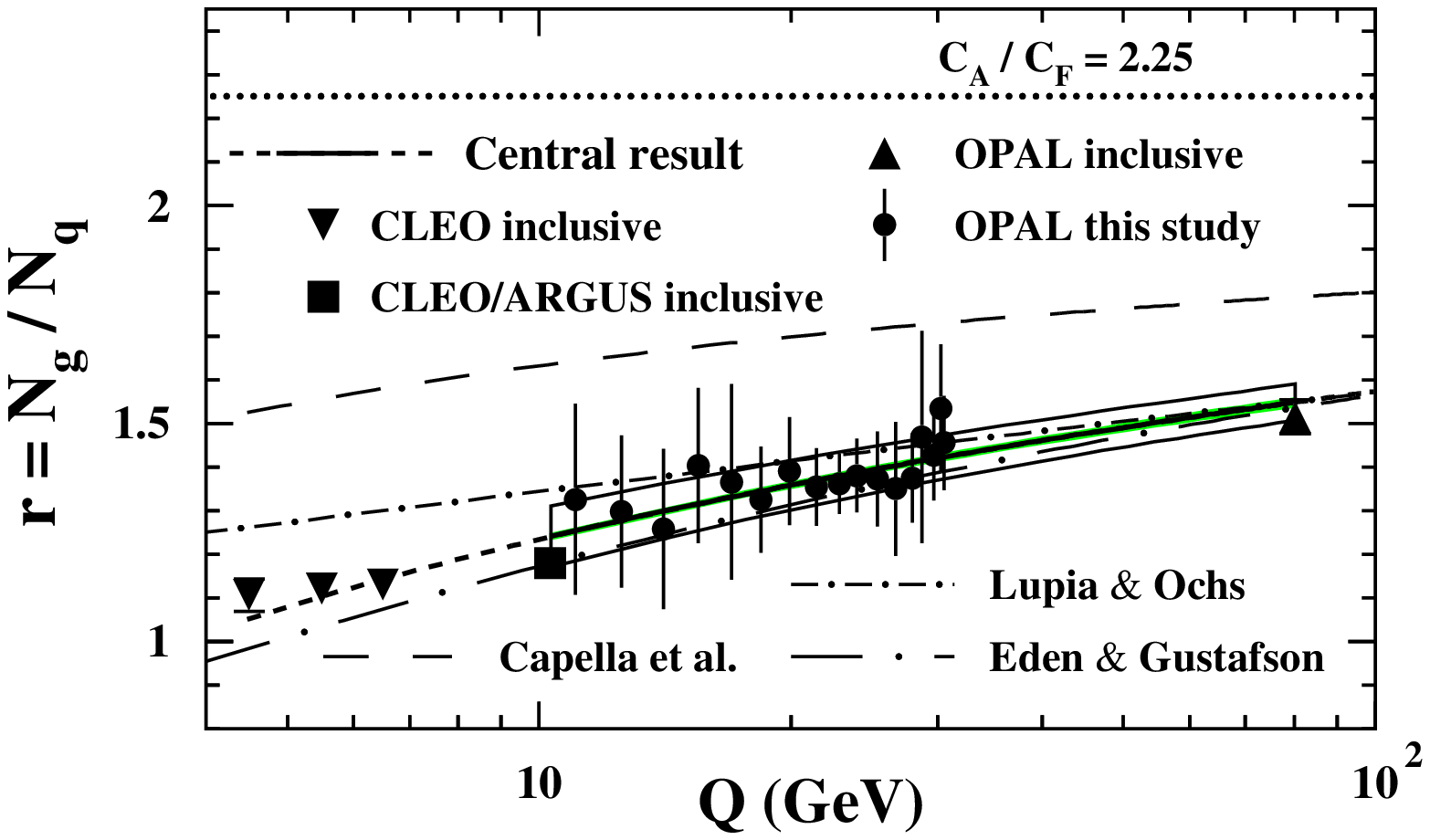} &
    \epsfxsize=8.5pc \epsffile{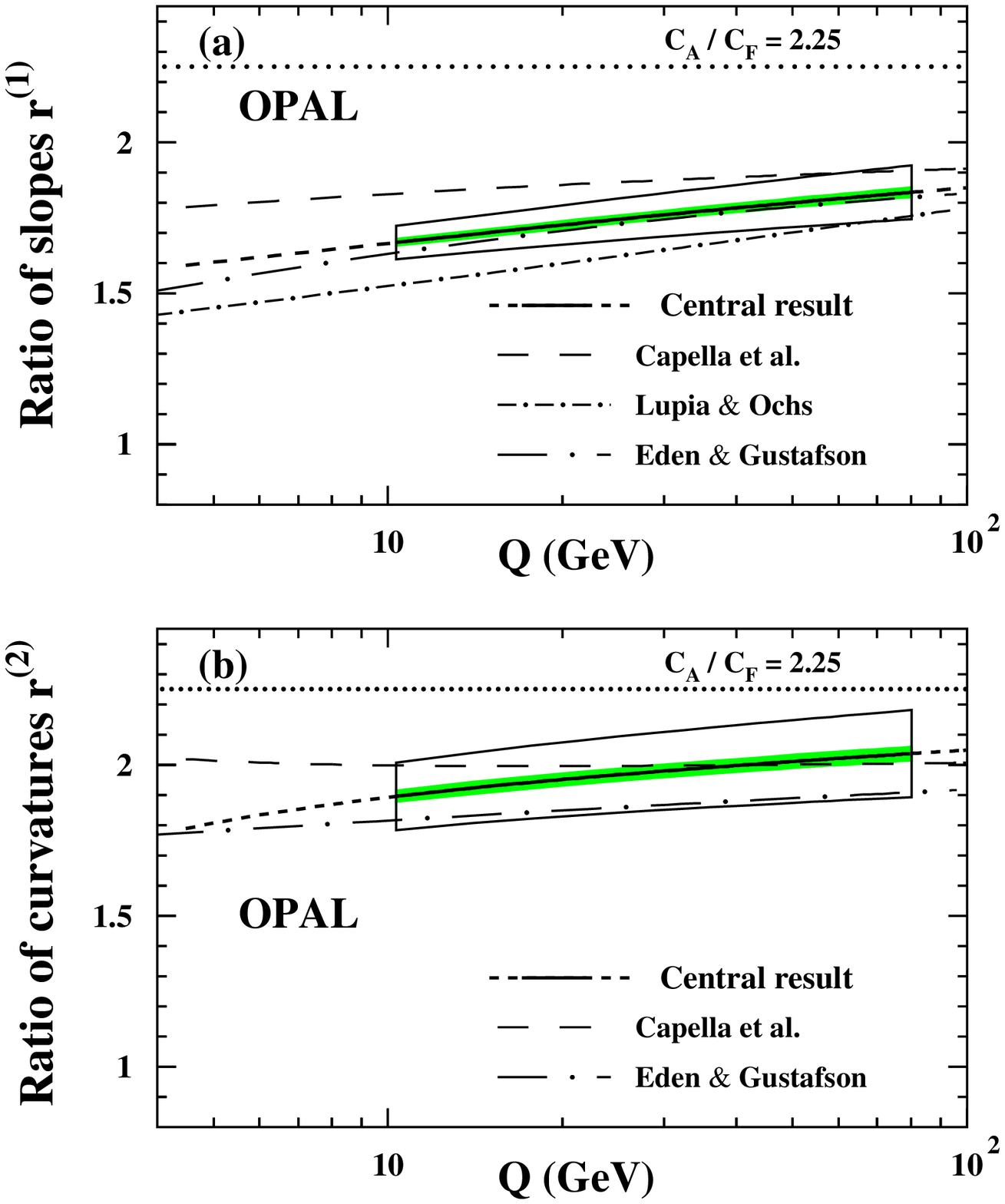} \\
  \end{tabular}
\end{center}
\caption{Leftmost two plots:
the mean charged particle multiplicity of unbiased
$gg$ events as a function of scale.
Central plot: 
the ratio of slopes $r$
between unbiased gluon and uds quark jets.
Rightmost two plots:
the corresponding results for the ratios of 
slopes $r^{(1)}$ and of curvatures~$r^{(2)}$.
}
\label{fig-ggmult}
\end{figure}

The leftmost set of plots in 
Fig.~\ref{fig-ggmult} shows our results for the unbiased 
charged particle multiplicities of $gg$ events, $\nggch$.
The solid points in the top plot
are obtained from eq.~(\ref{eq-eden14b})
using the direct measurements of~$\nqqch\,(L,\ktlu)$.
The asterisks and open symbols in the bottom plot
show the corresponding results from eqs.~(\ref{eq-eden14b}) 
and~(\ref{eq-eden14a}) using the calculated expressions for
$\nqqch\,(J,\ktlu)$ from eq.~(\ref{eq-qbiased}).
The two sets of results from eq.~(\ref{eq-eden14b})
(solid points in the top left plot
and asterisks in the bottom left plot) 
are seen to be very similar.

Included in the leftmost plots are direct measurements 
of unbiased gluon jet multiplicity
from the CLEO~\cite{bib-cleo92,bib-cleo97} 
and OPAL Collaborations~\cite{bib-opalgincl98}.
Also shown are the predictions of the
Herwig and Jetset Monte Carlo event generators for
the inclusive charged particle multiplicity of $gg$ events.
It is seen that the Monte Carlo
predictions describe the direct measurements of unbiased
gluon jet multiplicity (triangle symbols) well.
The results from the present analysis based on
eq.~(\ref{eq-eden14b}) 
are also well described by the Monte Carlo curves.
In contrast,
the results from eq.~(\ref{eq-eden14a}) 
(open symbols in the bottom leftmost plot) 
are generally well above the Monte Carlo predictions,
and --~if extrapolated to lower and higher energies~--
appear inconsistent with the direct
measurements from CLEO and OPAL as well.
We therefore conclude that the 
equation based on the Lund definition 
of the gluon jet scale, eq.~(\ref{eq-eden14b}),
yields results which are more consistent with other
studies than the equation based on the 
Leningrad definition, eq.~(\ref{eq-eden14a}).
We henceforth restrict our analysis of gluon jets to the former
set of results.

We employ the following procedure to determine
$r$, $r^{(1)}$ and~$r^{(2)}$ from experiment.
The corrected unbiased quark and gluon jet multiplicities
are separately fitted using the 
next-to-next-to-next-to-leading order
(3NLO) expressions for 
$\nq$ and $\ngluon$~\cite{bib-capella,bib-dgaryplb},
respectively.
To better constrain the results,
the direct measurements of unbiased gluon jet multiplicity
from CLEO and OPAL are included in the gluon jet fit.
The ratio of the fitted expressions for 
$\ngluon$ and $\nq$ defines~$r$.
We calculate the first and second derivitives of
the analytic equations for $\ngluon$ and $\nq$ with respect to $y$
and evaluate the resulting expressions using the corresponding
fitted parameter values.
The ratios of these terms define~$r^{(1)}$ and~$r^{(2)}$.

Our results for $r$ are shown in the central plot
in Fig.~\ref{fig-ggmult},
those for $r^{(1)}$ and~$r^{(2)}$ in 
the two rightmost plots.
The central results are indicated by solid curves.
The shaded bands show the statistical uncertainties.
The overall uncertainties,
with statistical and systematic terms added in quadrature,
are shown by the open bands.
At 30~GeV,
a typical scale in our analysis,
we find
$r$$\,=\,$$1.422\pm0.006\pm0.051$,
$r^{(1)}$$\,=\,$$1.761\pm0.013\pm0.070$ and
$r^{(2)}$$\,=\,$$1.98\pm0.02\pm0.13$.
This is consistent with the QCD prediction~\cite{bib-capella}
that $r$$\;<\;$$r^{(1)}$$\,<\,$$r^{(2)}$$\,<\,$$\caa$/$\cff$$\,=\,$2.25
for the scales accessible in our study,

The 3NLO predictions of Capella et al.~\cite{bib-capella} 
for $r$, $r^{(1)}$ and $r^{(2)}$
are shown by the long-dashed curves in 
the central and rightmost plots of Fig.~\ref{fig-ggmult}.
The analytic predictions for $r$ and $r^{(1)}$
exceed the corresponding
experimental results by about 22\% and~6\%,
while the theory agrees with the data for~$r^{(2)}$.
A second QCD prediction for $r$ versus scale was
recently presented by Lupia and Ochs~\cite{bib-lupia}.
The predictions of this calculation for $r$ 
and $r^{(1)}$ are shown by the short-dashed-dotted curves.
At 30~GeV,
$r$ is predicted to be~1.45,
in agreement with our measurement at that scale.
The corresponding result for $r^{(1)}$ is 1.64,
about 7\% below the data.
Finally,
theoretical predictions for $r$, $r^{(1)}$ and~$r^{(2)}$
can be derived from the formalism of
Ed\'{e}n and Gustafson~\cite{bib-eden}.
These predictions
are shown by the long-dash-dotted curves.
The results are seen to be in good 
overall agreement with the data.
We note, however,
that these predictions are based on the
experimental measurements of quark jet multiplicities.
Therefore,
the predictions we derive based on~\cite{bib-eden}
are not entirely independent of the data.

In conclusion,
we find overall agreement between the
experimental and theoretical results.


\begin{thebibliography}{99}
\bibitem{bib-cleo92}
  CLEO Collaboration, M.S. Alam et al., Phys. Rev. {\bf D46} (1992)~4822.
\bibitem{bib-cleo97}
  CLEO Collaboration, M.S. Alam et al., Phys. Rev. {\bf D56} (1997) 17.
\bibitem{bib-opalgincl96}
  OPAL Collaboration, G. Alexander et al., Phys. Lett. {\bf B388} (1996)~659.
\bibitem{bib-opalgincl97}
  OPAL Collaboration, K. Ackerstaff et al., Eur. Phys. J. {\bf C1} (1998)~479.
\bibitem{bib-opalgincl98}
  OPAL Collaboration, G. Abbiendi et al., Eur. Phys. J. {\bf C11} (1999)~217.
\bibitem{bib-lupia}
  S. Lupia and W. Ochs, Phys. Lett. {\bf B418} (1998) 214.
\bibitem{bib-capella}
  A. Capella et al., Phys. Rev. {\bf D61} (2000) 074009.
\bibitem{bib-eden}
  P. Ed\'{e}n and G. Gustafson, JHEP {\bf 9809} (1998) 015.
\bibitem{bib-edenkhoze}
  P. Ed\'{e}n, G. Gustafson and V. Khoze, Eur. Phys. J. {\bf C11} (1999)~345.
\bibitem{bib-dok88}
  Yu.L. Dokshitzer, V.A. Khoze and S.I. Troyan,
  Sov. J. Nucl. Phys. {\bf 47} (1988) 881.
\bibitem{bib-delphiprelim}
  DELPHI Collaboration, Conference notes 2000-118 CONF 417
  and 2001-067 CONF 495
\bibitem{bib-opalse91}
  OPAL Collaboration, M.Z. Akrawy et al., 
  Phys. Lett. {\bf B261} (1991)~334.
\bibitem{bib-opalqg91}
  OPAL Collaboration, G. Alexander et al.,
  Phys. Lett. {\bf B265} (1991)~462.
\bibitem{bib-dgaryplb}
  I.M. Dremin and J.W. Gary, Phys. Lett. {\bf B459} (1999)~341.
\end{thebibliography}
\end{document}